\documentclass[aps,prl,twocolumn,superscriptaddress,amssymb,floatfix]{revtex4}
\usepackage{graphicx}
\usepackage{epstopdf}
\usepackage[T1]{fontenc}
\usepackage[latin9]{inputenc}
\usepackage{amsbsy}

\newcommand{\sigvec}[0]{\mbox{\boldmath $\sigma$}}

\newcommand{\jqm}[1]{#1}
\newcommand{\jqd}[1]{}
\newcommand{\jqc}[1]{}
\usepackage{color}
\begin{document}


\title{
Non-unitary triplet pairing in the centrosymmetric superconductor LaNiGa$_2$
}


\author{A.D. Hillier}
\affiliation{ISIS facility, STFC Rutherford Appleton Laboratory, Harwell Science and Innovation Campus, Oxfordshire, OX11 0QX, UK}
\author{J. Quintanilla}
\affiliation{ISIS facility, STFC Rutherford Appleton Laboratory, Harwell Science and Innovation Campus, Oxfordshire, OX11 0QX, UK}
\affiliation{SEPnet and Hubbard Theory Consortium, School of Physical Sciences, University of Kent, Canterbury CT2 7NH, UK}
\author{B. Mazidian}
\affiliation{ISIS facility, STFC Rutherford Appleton Laboratory, Harwell Science and Innovation Campus, Oxfordshire, OX11 0QX, UK}
\affiliation{H. H. Wills Physics Laboratory, University of Bristol, Tyndall Avenue, Bristol BS8 1TL, UK}

\author{J. F. Annett}
\affiliation{H. H. Wills Physics Laboratory, University of Bristol, Tyndall Avenue, Bristol BS8 1TL, UK}
\author{R. Cywinski}
\affiliation{School of Applied Sciences, University of Huddersfield, Queensgate, Huddersfield, HD1 3DH, UK}

\date{\today}

\begin{abstract}
Muon spin rotation and relaxation experiments on the centrosymmetric intermetallic superconductor LaNiGa$_2$ are reported. The appearance of spontaneous magnetic fields coincides with the onset of superconductivity, implying that the superconducting state breaks  time reversal symmetry, similarly to non-centrosymmetric LaNiC$_2$.  Only four triplet states are  compatible with this observation, all of which are non-unitary triplets.  This suggests that LaNiGa$_2$ is the centrosymmetric analogue of LaNiC$_2$. 
We argue that these materials are representatives of  a new family of paramagnetic non-unitary superconductors.
\end{abstract}


\maketitle


Symmetry breaking is a central concept of physics for which superconductivity provides one of the best understood  paradigms. In a conventional superconductor \cite{Bardeen57_2} \jqc{Added space before all `cite' commands.} gauge symmetry is broken, while unconventional superfluids and superconductors break other symmetries as well \cite{Sigrist91}. Examples include $^3$He \cite{Lee97}, cuprate high-temperature superconductors \cite{Annett90}, the ruthenate Sr$_2$RuO$_4$ \cite{Mackenzie03} and
more recently,
non-centrosymmetric LaNiC$_{2}$ \cite{Hillier09}.  
The latter has weak spin-orbit coupling (SOC) \cite{Quintanilla10}, low symmetry and 
is a non-unitary superconductor. In a non-unitary superconductor the pairing states of the spin-up and spin-down Fermi surfaces are different. At the instability, a spin-up superfluid can coexist with spin-down Fermi liquid. While non-unitary triplet superconductivity is well-established in ferromagnetic superconductors \cite{2010-Visser}, its occurrence in parmagnetic LaNiC$_2$ remains puzzling. Here we provide experimental evidence of this phenomenon in another, compositionally related, but centrosymmetric superconductor: LaNiGa$_2$. We also advance an explanation in terms of a coupling between triplet instabilities and paramagnetism that is quite generic and for which these two could provide the first examples of what might be a larger class of materials.

In general unconventional pairing can be difficult to establish  in any given material. However evidence for  time-reversal symmetry (TRS) breaking 
in particular can be shown through the detection of spontaneous but very small internal fields
 \cite{Sigrist91}. Muon spin relaxation/rotation ($\mu$SR) is especially sensitive for detecting small changes in internal fields and can easily measure fields of 0.1~G which corresponds to $\approx$0.01~$\mu_B$. This makes $\mu$SR an extremely powerful technique for measuring the effects of TRS breaking in exotic superconductors. Direct observation of TRS breaking states is extremely rare and spontaneous fields have been observed in this way only in a few systems: PrOs$_4$Sb$_{12}$ \cite{Aoki03}, Sr$_2$RuO$_4$ \cite{Luke98} (where TRS breaking was subsequently confirmed by optical measurements \cite{Xia06}), B-phase of UPt$_3$ \cite{Luke93} (although not without controversy  \cite{deReotier95,Higemoto00}), (U,Th)Be$_{13}$ \cite{Heffner90} and more recently LaNiC$_2$ \cite{Hillier09}, PrPt$_4$Ge$_{12}$ \cite{Maisuradze10} and (PrLa)(OsRu)$_4$Sb$_{12}$ \cite{Shu11}. 
For examples of other systems where the effect is \textit{not} observed see 
Refs.~\cite{Adroja05,Anand11,Tran10,Hillier11}.  
 Broken TRS in superconductors  is especially interesting, because it implies not just unconventional pairing,  but the existence of  two-fold or higher degeneracy of the superconducting order parameter space \cite{Mineev99}. 

The observation of broken TRS in LaNiC$_2$ was particularly surprising because of the low symmetry of this orthorhombic, non-centrosymmetric, material\cite{Hillier09}.  
Symmetry analysis has shown that the low dimensionality of this structure, with C$_{2v}$ point group, gives rise to only 12 possible gap functions. Of these only 4 break TRS and these are all non-unitary triplet pairing states \cite{Hillier09}. These four gap functions are all derived from one dimensional irreducible
representations of the point group, implying that the only possible order parameter degeneracy is derived from the triplet Cooper pair spin orientational
 degree of freedom.   A subsequent analysis of the effects of SOC on this system \cite{Quintanilla10} shows that the SOC always lifts this final degeneracy,
leading to a completely non-degenerate order parameter  space, which would not be expected to allow spontaneous breaking of TRS at the 
superconducting transition temperature, $T_c$ \cite{Mineev99}.  The only way to reconcile this with the experimental observations of broken TRS in this material is to assume that the effect of SOC is weak on the relevant electron states at the Fermi level in this material. Additional experimental evidence for unconventional pairing in LaNiC$_2$ has been reported recently \cite{Bonalde11}.

\begin{figure}
\includegraphics[width=9cm,clip=true,trim= 60 120 30 81]{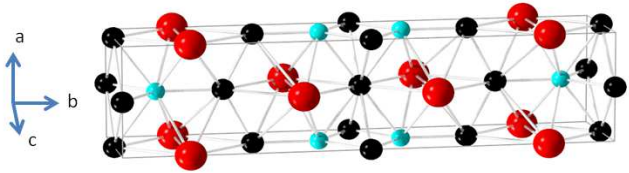}
\caption{\label{Fig:Struct} (color online) The orthorhombic crystal structure of LaNiGa$_{2}$. The red spheres (largest) are La, blue spheres (smallest) are Ni and the black spheres (medium) are Ga.}
\end{figure}

In this letter we report $\mu$SR results on the centrosymmetric superconductor LaNiGa$_2$ showing that TRS is broken on entering the superconducting state.
This is a centrosymmetric material, which crystallises in the NdNiGa$_{2}$ orthorhombic structure, with space group Cmmm (D$_{2h}$)   \cite{Zeng02} (see Fig. \ref{Fig:Struct}).  Magnetisation and heat capacity measurements have previously shown that LaNiGa$_{2}$ is a paramagnetic superconductor, with a T$_c$ onset of 2.1~K\cite{Zeng02}. 
 Heat capacity measurements have shown a specific heat jump $\Delta C / \gamma T_c \approx$1.31\footnote{The authors of Ref \cite{Zeng02} have determined a higher value of $\Delta C / \gamma T_c$ by using a 10 \% - 90\% definition of $T_c$}, which is slightly lower than the expected BCS value of 1.43, and the temperature dependence has the conventional BCS exponential form. Specific heat jumps in TRS-breaking superconductors are sometimes lower and sometimes higher than the BCS value, for example Sr$_2$RuO$_4$ \cite{Nishizaki00}, LaNiC$_2$\cite{Pecharsky98} and PrPt$_4$Ge$_{12}$\cite{Gumeniuk08}, respectively. 
 Below we analyse the possible order parameter symmetries. 
Although LaNiGa$_2$ has a different point group to LaNiC$_2$ and is centrosymmetric,  
it is similar to LaNiC$_2$ in that it has
only 12 possible superconducting  symmetries \cite{Annett90}. For both materials these are all 
derived from one-dimensional irreducible representations. In particular, only four of these 
states break TRS and these are all non-untiary triplet pairing states (see Table \ref{tab:states}a).

The sample was prepared by melting together stoichiometric amounts of the constituent elements in a water-cooled argon arc furnace.  The $\mu$SR experiments were carried out using the MuSR spectrometer in longitudinal and transverse geometries. At the ISIS facility, a pulse of muons is produced every 20 ms and has a FWHM of $\sim$70 ns. These muons are implanted into the sample and decay with a half-life of 2.2~$\mu$s into a positron which is emitted preferentially in the direction of the muon spin axis and two neutrinos. These positrons are detected and time stamped in the detectors which are positioned either before, F, or after, B, the sample for longitudinal (relaxation) experiments. The forward and backward detectors are each segmented into 32 detectors. Using these counts the asymmetry in the positron emission can be determined and, therefore, the muon polarisation is measured as a function of time. For the transverse field experiments, the magnetic field was applied perpendicular to the initial muon spin direction and momentum. For a more detailed description of the different instrumental geometry can be found in Ref. \cite{blackbook, Yaouanc, Schenck}.

The sample was powder mounted onto a 99.995+~$\%$ pure silver plate. Any muons stopped in silver give a time independent background for longitudinal (relaxation) experiments. The sample holder and sample were mounted onto a TBT dilution refrigerator with a temperature range of 0.045-4~K. The stray fields at the sample position are cancelled to within 1~$\mu$T by a flux-gate magnetometer and an active compensation system controlling the three pairs of correction coils. The transverse field $\mu$SR (TF-$\mu$SR) experiment was conducted with applied fields between 5~mT and 60~mT, which ensured the sample was in the mixed state. Each field was either applied above the superconducting transition and the sample was then cooled to base temperature (FC) or the sample was first cooled to base temperature and then the field was applied (ZFC). The sample was cooled to base temperature in zero field and the $\mu$SR spectra were collected upon warming the sample while still in zero field. 

\begin{figure}[tbh]
\includegraphics[width=7.0cm]{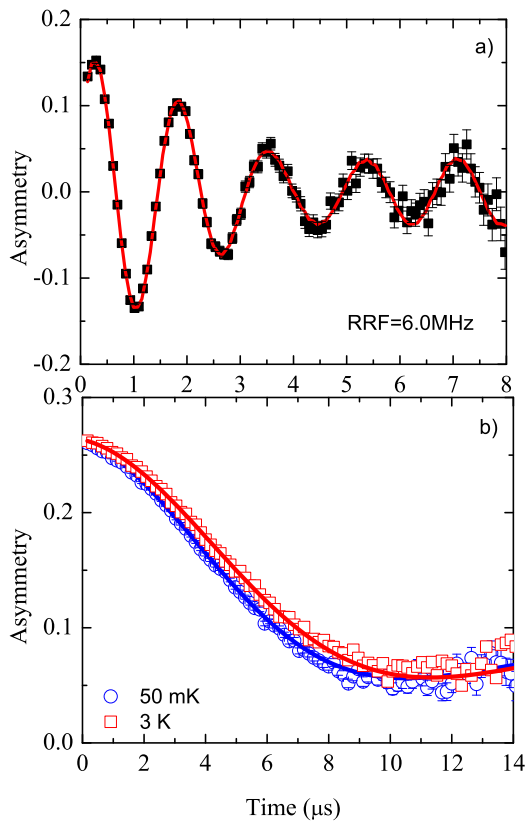}
\caption{\label{fig:spectra}(Color online) The upper graph is a typical muon asymmetry spectra in LaNiGa$_2$ taken in a transverse field of 40 mT at 0.05 K (shown in the rotating reference frame (RRF) of 6.0~MHz. The line is a fit to the data using Eqn. \ref{eqn:fitfunctf}. For clarity, only one of the two virtual detectors have been shown. The lower graph is the zero field $\mu$SR spectra for LaNiGa$_2$.  The blue symbols are the data collected at 56~mK and the red symbols are the data collected at 3.0~K. The lines are a least squares fit to the data.}
\end{figure}

The MuSR spectrometer comprises 64 detectors. 
In software, the data is mapped to two orthogonal virtual detectors each characterised by a phase offset $\varphi$.
The resulting 2 spectra were simultaneously fitted with a sinusoidal oscillating function with Gaussian relaxation:
\begin{equation}
G_z(t) = \displaystyle\sum_{i=1}^2 A_i exp(-\frac{\sigma_i ^2t^2}{2})cos(2\pi \nu_i t + \varphi)
\label{eqn:fitfunctf}
\end{equation}
\noindent 
where the \emph{i}th index denotes the sample and background contributions, respectively, $A_i$ is the initial asymmetry, $\sigma_i$ is the Gaussian relaxation rate and $\nu_i$ is the muon spin precessional frequency. The background term comes from those muons which were implanted into the silver sample holder and therefore this oscillating term has no depolarisation, i.e. $\sigma_2$=0.0~$\mu s^{-1}$, as silver has a negligible nuclear moment. Fig. \ref{fig:spectra} shows a typical spectrum for LaNiGa$_2$ with an applied field of 40~mT at 50~mK after being FC. 

\begin{figure}[tbh]
\includegraphics[width=7.cm]{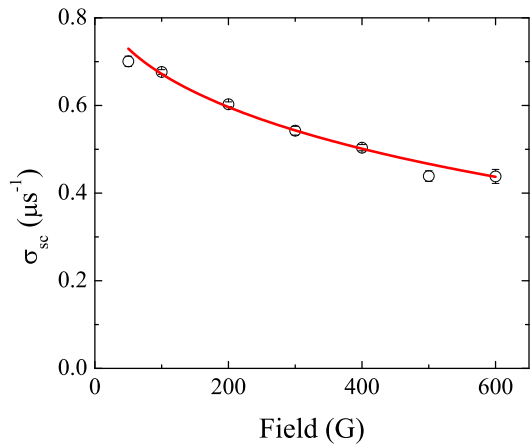}
\caption{\label{fig:SvB}(Color online) The field dependence of $\sigma$ at 50mK, after being field cooled. The line is a fit to the data using Eqn. \ref{eqn:Brandt}.}
\end{figure}

\begin{figure}
\includegraphics[width=8.5cm]{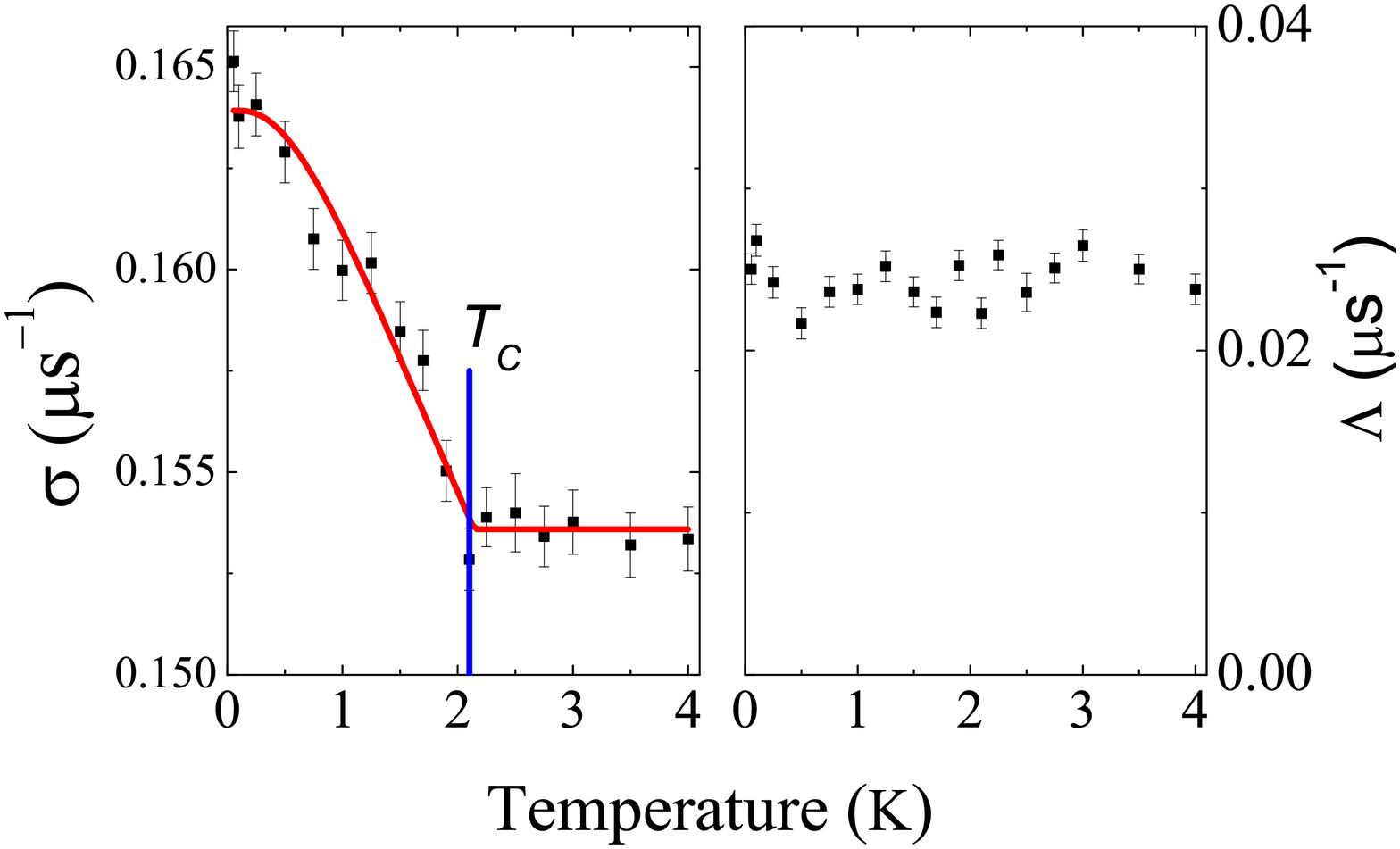}
\caption{\label{fig:SigmavT} (color online) The left graph shows the temperature dependence of  $\sigma$, for LaNiGa$_2$ in zero-field, which clearly shows the spontaneous fields appearing at T$_c$=2.1~K (shown has the vertical line). The line is fit to the data using an approximation\cite{Carrington03} to the BCS order parameter for $\sigma_e$. The right graph shows the temperature dependence of the electronic relaxation rate, $\Lambda$, for LaNiGa$_2$ in zero-field, which  shows no temperature dependence.} 
\end{figure}

As the muon spin rotation arising from the field distributions associated with the flux line lattice is independent of that arising from the nuclear moments we can write $\sigma_{1}^{2}=\sigma_{sc}^{2}+\sigma_{n}^{2}$. $\sigma_{n}$ is assumed to be constant in this temperature region, and is determined from measurements just above $T_{c}$. Each data point was collected after field cooling the sample from above $T_c$. The field dependence of $\sigma_{sc}$ (in $\mu$s) is related to the superconducting penetration depth $\lambda$ (in nm) and coherence length $\xi$ via the relation 
\begin{equation}
\sigma_{sc}=4.83\times 10^4(1-b)(1+1.21(1-\sqrt{b})^3)\lambda^{-2}
\label{eqn:Brandt}
\end{equation}      

\noindent where \jqm{$b=B/B_{C2}$ is} the ratio of applied field to upper critical field. From this we have determined $\lambda$ and B$_{C2}$ and hence $\xi$ to be 350(10)~nm, 410(3)~mT and 28(3)~nm respectively (see Fig. \ref{fig:SvB}). This shows that LaNiGa$_2$ is a type II superconductor, with a superconducting electron density and effective superelectron mass of $9\times10^{26}~m^{-3}$ and $3.9~m_e$\jqm{, respectively}. More details on these calculations can be found in Ref  \cite{Hillier97}. 
Now let us consider the longitudinal $\mu$SR data. The absence of a  precessional signal in the $\mu$SR spectra at all temperatures confirms that there are no spontaneous coherent internal magnetic fields associated with long range magnetic order in LaNiGa$_2$ at any temperature. In the absence of atomic moments muon spin relaxation is expected to arise entirely from the local fields associated with the nuclear moments. These nuclear spins are static, on the time scale of the muon precession, and are randomly orientated. The depolarisation function, $G_z(t)$, can be described by the Kubo-Toyabe function \cite{Hayano79}
\begin{equation}
\label{eq:fitfun}
G_z^{KT}(t)=(\frac{1}{3}+\frac{2}{3}(1-\sigma^2 t^2)\exp(-\frac{\sigma^2t^2}{2})),
\end{equation}
\noindent where $\sigma/\gamma_\mu$ is the local field distribution width and  $\gamma_\mu=2\pi \times 135.5$ MHz T$^{-1}$ is the muon gyromagnetic ratio. The spectra that we observed for LaNiGa$_2$ are well described by the function
\begin{equation}
\label{eq:fitfun2}
G_z(t)=A_0G_z^{KT}(t)\exp(-\Lambda t) + \\
A_{bckgrd},
\end{equation}
\noindent where $A_0$ is the initial asymmetry, A$_{bckgrd}$ is the background, and $\Lambda$ is the electronic relaxation rate (see Fig. \ref{fig:spectra}). It is assumed that the exponential factor involving $\Lambda$ arises from electronic moments which afford an entirely independent muon spin relaxation channel in real time.
%
%
The only parameter that shows any temperature dependence is $\sigma$, which increases rapidly with decreasing temperature below T$_c$ (see Fig. \ref{fig:SigmavT}). We interpret this increase in $\sigma$ as a signature of a coherent internal field with a very low frequency as discussed by Aoki {\it et al.}
 \cite{Aoki03} for PrOs$_4$Sb$_{12}$. This increase in $\sigma$ has been modelled, assuming that there are uncorrelated, by $\sigma(T)^2=\sigma_n^2+\sigma_e(T)^2$, where $\sigma_n$ and $\sigma_e$ are the nuclear and electronic contributions respectively. The temperature dependence of $\sigma_e$ agrees with the BCS order parameter (see Fig. \ref{fig:SigmavT}).
 
%
%
Let us now discuss the implications of this result for the pairing symmetry. The group theoretical analysis 
for the D$_{2h}$ point group of this system has already been investigated  \cite{Annett90}. For the simplest case, where translational symmetry is not broken and SOC does not play a role, this point group has a total of $8$ irreducible representations. This leads to $12$ possible order parameters, as given in Table \ref{tab:states}a. Of these $12$, $8$ are unitary and $4$ are non-unitary. Only the $4$ non-unitary order parameters have a non-trivially complex order parameter that can break TRS. In the case where SOC is large there are only $4$ possible states
(see Table \ref{tab:states}b) and none of them break TRS. Therefore, like LaNiC$_2$, LaNiGa$_2$ must be a non-unitary triplet superconductor with weak SOC. As we have predicted for LaNiC$_{2}$\cite{Quintanilla10} if the SOC is not zero then a split transition would be expected.

\begin{table}
(a)
  \begin{tabular}{|c|c|c|}
    \hline
    \textbf{$SO(3) \times D_{2h}$}
    & 
    \begin{tabular}{c}
      gap function
      \tabularnewline
      (unitary)
    \end{tabular}
    &
    \begin{tabular}{c}
      gap function
      \tabularnewline
      (non-unitary)
    \end{tabular}
    \tabularnewline
    \hline
    \hline
    $^1A_{1}$
    &
    $\Delta({\bf k}) = 1$
    &
    -
    \tabularnewline
    $^1B_{1}$
    &
    $\Delta({\bf k}) = XY$
    &
    -
    \tabularnewline
    $^1B_{2}$
    &
    $\Delta({\bf k}) = XZ$
    &
    -
    \tabularnewline
    $^1B_{3}$
    &
    $\Delta({\bf k}) = YZ$
    &
    -
    \tabularnewline
    \hline  
    $^3A_{1}$
    &
    ${\bf d}({\bf k}) = (0,0,1)XYZ$
    &
    ${\bf d}({\bf k}) = (1,i,0)XYZ$
    \tabularnewline
    $^3B_{1}$
    &
    ${\bf d}({\bf k}) = (0,0,1)Z$
    &
    ${\bf d}({\bf k}) = (1,i,0)Z$
    \tabularnewline
    $^3B_{2}$
    &
    ${\bf d}({\bf k}) = (0,0,1)Y$
    &
    ${\bf d}({\bf k}) = (1,i,0)Y$
    \tabularnewline
    $^3B_{3}$
    &
    ${\bf d}({\bf k}) = (0,0,1)X$
    &
    ${\bf d}({\bf k}) = (1,i,0)X$
    \tabularnewline
    \hline
  \end{tabular}
  
(b)
  \begin{tabular}{|c|c|}

    \hline

   \textbf{$D_{2h}$}
  & 
      gap function with strong SOC

    \tabularnewline
  
    \hline  
    $A_{1}$
    &
    ${\bf d}({\bf k}) = (AX,BY,CZ)$
   
    \tabularnewline
    $B_{1}$
    &
    ${\bf d}({\bf k}) = (AY,BX,CXYZ)$
  
    \tabularnewline
    $B_{2}$
    &
    ${\bf d}({\bf k}) = (AZ,BXYZ,CX)$
   
    \tabularnewline
    $B_{3}$
    &
    ${\bf d}({\bf k}) = (AXYZ,BZ,CY)$

    \tabularnewline
    \hline
  \end{tabular}
\caption{\label{tab:states} The upper table (a) shows the gap function of the homogeneous superconducting states allowed by symmetry, for weak spin-orbit coupling. We have used the standard notation $\hat{\Delta}({\bf k})=\Delta({\bf k})i\hat{\sigma}_y$ for 
singlet states and $\hat{\Delta}({\bf k})=i\left[{\bf d}({\bf k}).\hat{\sigvec}\right]\hat{\sigma}_y$ for 
triplets, where $\hat{\sigvec}\equiv (\hat{\sigma}_x,\hat{\sigma}_y,\hat{\sigma}_z)$ is the vector of Pauli 
matrices.  Here $X$, $Y$ and $Z$ are  any basis functions
in the Brillouin zone of odd parity, such as $\sin{k_x}$,  $\sin{k_y}$ and  $\sin{k_z}$, respectively.
Only the four non-unitary triplet states are compatible with our observation of broken TRS. The lower table (b) shows the homogeneous spin-triplet superconducting states 
 allowed by symmetry at $T_c$, for the case of strong spin-orbit coupling.  
The spin singlet gap functions are the same as in the weak spin-orbit case.
The notation of the possible gap functions is the same as the upper table, except that here
$A$, $B$ and $C$  are (real) constants determined by the microscopic gap equation. 
 Clearly none of these states breaks TRS at $T_c$.}
\end{table}

Until the discovery of non-unitary triplet pairing in LaNiC$_{2}$
this state had only been confirmed in ferromagnetic superconductors \cite{2010-Visser}.
The additional observation of non-unitary triplet pairing in LaNiGa$_{2}$
brings up the question of how a triplet superconductor whose normal
state is paramagnetic could favour this state. The usual Landau free energy describing a triplet pairing
instability in our system is of the form 
\begin{equation}
F=a\left|\boldsymbol{\eta}\right|^{2}+\frac{b}{2}\left|\boldsymbol{\eta}\right|^{4}+b'\left|\boldsymbol{\eta \times \eta^{*}}\right|^{2}
\label{F_traditional}
\end{equation}
where $\boldsymbol{\eta}$ is the order parameter, which relates to
the $\mathbf{d}$ vector through 
$\mathbf{d}\left(\mathbf{k}\right)=\boldsymbol{\eta}\Gamma\left(\mathbf{k}\right)$
{[}the possible functional forms of $\Gamma\left(\mathbf{k}\right)$
are given in Table \ref{tab:states}a{]}.
The triplet instability takes place
when $a=0$, which determines $T_{c}$ and is independent of whether
pairing is unitary or non-unitary. Below $T_{c}$, the second of the quartic terms
decides which of the two states is most stable. 
The criterion for non-unitary triplet pairing is  \cite{Annett90} 
\begin{equation}
b'<0.
\label{nouni-condition}
\end{equation} On the other hand
for a paramagnet there must be an additional term 
coupling
$\boldsymbol{\eta}$
to the magnetization $\mathbf{m}$. On symmetry grounds the simplest form of the free energy that takes this into account is 
\begin{equation}
F=a\left|\boldsymbol{\eta}\right|^{2}+\frac{\boldsymbol{m}^{2}}{2\chi}+\frac{b}{2}\left|\boldsymbol{\eta}\right|^{4}+b'\left|\boldsymbol{\eta \times \eta^{*}}\right|^{2}+b''\mathbf{m}\cdot\left(i\boldsymbol{\eta}\times\boldsymbol{\eta}^{*}\right).
\label{F_full}
\end{equation}
Here $\chi$ is the normal state susceptibility. 

For given $\boldsymbol{\eta}$ the last term on the right hand side of (\ref{F_full}) describes an effective magnetic field $\mathbf{h}_{eff}=-b''\left(i\boldsymbol{\eta}\times\boldsymbol{\eta}^{*}\right)$ coupled to $\boldsymbol{m}$. This field vanishes for unitary triplet pairing, but in the non-unitary case it induces a magnetization
\begin{equation}
\mathbf{m}=- \chi b'' \left(i\boldsymbol{\eta}\times\boldsymbol{\eta}^{*}\right).
\label{m_value}
\end{equation}
Below $T_c$, $\eta \sim \left(T_c-T\right)^{1/2}$ whence $m \sim T_c-T$. This subdominant order parameter lowers the energy of the non-unitary state compared to the unitary one. Indeed substituting (\ref{m_value}) into (\ref{F_full}) we recover the simpler expression (\ref{F_traditional}) but with the $b'$ coefficient replaced with $b'-b''^2 \chi/2$. The condition (\ref{nouni-condition}) 
then becomes 
\begin{equation}
b'-{b''}^{2}\chi/2<0.
\end{equation}
 For a paramagnet the second term on the left hand side is
always negative, favouring non-unitary triplet pairing states. 
This effect would be expected to be strongest in proximity to a Stoner instability. We note that in superconducting ferromagnets \cite{2010-Visser}
the same coupling term exists and stabilizes non-unitary triplet pairing states
by increasing their $T_c$ relative to unitary states.

In conclusion, zero field and transverse field $\mu$SR experiments have been carried out on LaNiGa$_2$. The zero field measurements show a spontaneous field appearing at the superconducting transition temperature. This provides convincing evidence that time reversal symmetry is broken in the superconducting state of 
LaNiGa$_2$. Symmetry analysis implies non-unitary triplet pairing, in close analogy with the non-centrosymmetric superconductor LaNiC$_2$. We propose that these materials could represent a new class of superconductors where a triplet superconducting instability of a paramagnetic state gives rise to non-unitary pairing through a generic coupling to the magnetization.

\begin{acknowledgments}
This works was supported by EPSRC and STFC (U.K.). J.Q. gratefully acknowledges funding from HEFCE and STFC through the South-East Physics network (SEPnet).
\end{acknowledgments}

\bibliography{LaNiGa2_prl_ver4}

\begin{thebibliography}{33}
\expandafter\ifx\csname natexlab\endcsname\relax\def\natexlab#1{#1}\fi
\expandafter\ifx\csname bibnamefont\endcsname\relax
  \def\bibnamefont#1{#1}\fi
\expandafter\ifx\csname bibfnamefont\endcsname\relax
  \def\bibfnamefont#1{#1}\fi
\expandafter\ifx\csname citenamefont\endcsname\relax
  \def\citenamefont#1{#1}\fi
\expandafter\ifx\csname url\endcsname\relax
  \def\url#1{\texttt{#1}}\fi
\expandafter\ifx\csname urlprefix\endcsname\relax\def\urlprefix{URL }\fi
\providecommand{\bibinfo}[2]{#2}
\providecommand{\eprint}[2][]{\url{#2}}

\bibitem[{\citenamefont{Bardeen et~al.}(1957)\citenamefont{Bardeen, Cooper, and
  Schrieffer}}]{Bardeen57_2}
\bibinfo{author}{\bibfnamefont{J.}~\bibnamefont{Bardeen}},
  \bibinfo{author}{\bibfnamefont{L.~N.} \bibnamefont{Cooper}},
  \bibnamefont{and} \bibinfo{author}{\bibfnamefont{J.~R.}
  \bibnamefont{Schrieffer}}, \bibinfo{journal}{Phys. Rev.}
  \textbf{\bibinfo{volume}{108}}, \bibinfo{pages}{1175} (\bibinfo{year}{1957}).

\bibitem[{\citenamefont{Sigrist and Ueda}(1991)}]{Sigrist91}
\bibinfo{author}{\bibfnamefont{M.}~\bibnamefont{Sigrist}} \bibnamefont{and}
  \bibinfo{author}{\bibfnamefont{K.}~\bibnamefont{Ueda}},
  \bibinfo{journal}{Rev. Mod. Phys.} \textbf{\bibinfo{volume}{63}},
  \bibinfo{pages}{239} (\bibinfo{year}{1991}).

\bibitem[{\citenamefont{Lee}(1997)}]{Lee97}
\bibinfo{author}{\bibfnamefont{D.~M.} \bibnamefont{Lee}},
  \bibinfo{journal}{Rev. Mod. Phys.} \textbf{\bibinfo{volume}{69}},
  \bibinfo{pages}{645} (\bibinfo{year}{1997}).

\bibitem[{\citenamefont{Annett}(1990)}]{Annett90}
\bibinfo{author}{\bibfnamefont{J.~F.} \bibnamefont{Annett}},
  \bibinfo{journal}{Adv. in Phys.} \textbf{\bibinfo{volume}{39}},
  \bibinfo{pages}{83} (\bibinfo{year}{1990}).

\bibitem[{\citenamefont{Mackenzie and Maeno}(2003)}]{Mackenzie03}
\bibinfo{author}{\bibfnamefont{A.~P.} \bibnamefont{Mackenzie}}
  \bibnamefont{and} \bibinfo{author}{\bibfnamefont{Y.}~\bibnamefont{Maeno}},
  \bibinfo{journal}{Rev. Mod. Phys.} \textbf{\bibinfo{volume}{75}},
  \bibinfo{pages}{657} (\bibinfo{year}{2003}).

\bibitem[{\citenamefont{Hillier et~al.}(2009)\citenamefont{Hillier,
  Quintanilla, and Cywinski}}]{Hillier09}
\bibinfo{author}{\bibfnamefont{A.~D.} \bibnamefont{Hillier}},
  \bibinfo{author}{\bibfnamefont{J.}~\bibnamefont{Quintanilla}},
  \bibnamefont{and} \bibinfo{author}{\bibfnamefont{R.}~\bibnamefont{Cywinski}},
  \bibinfo{journal}{Phys. Rev. Lett.} \textbf{\bibinfo{volume}{102}},
  \bibinfo{pages}{117007} (\bibinfo{year}{2009}).

\bibitem[{\citenamefont{Quintanilla et~al.}(2010)\citenamefont{Quintanilla,
  Hillier, Annett, and Cywinski}}]{Quintanilla10}
\bibinfo{author}{\bibfnamefont{J.}~\bibnamefont{Quintanilla}},
  \bibinfo{author}{\bibfnamefont{A.~D.} \bibnamefont{Hillier}},
  \bibinfo{author}{\bibfnamefont{J.~F.} \bibnamefont{Annett}},
  \bibnamefont{and} \bibinfo{author}{\bibfnamefont{R.}~\bibnamefont{Cywinski}},
  \bibinfo{journal}{Phys. Rev. B} \textbf{\bibinfo{volume}{82}},
  \bibinfo{pages}{174511} (\bibinfo{year}{2010}).

\bibitem[{\citenamefont{de~Visser}(2010)}]{2010-Visser}
\bibinfo{author}{\bibfnamefont{A.}~\bibnamefont{de~Visser}},
  \emph{\bibinfo{title}{Encyclopedia of Materials: Science and\ Tech\-nology}}
  (\bibinfo{publisher}{Elsevier, Oxford}, \bibinfo{year}{2010}), chap.
  \bibinfo{chapter}{Superconducting ferromagnets}.

\bibitem[{\citenamefont{Aoki et~al.}(2003)\citenamefont{Aoki, Tsuchiya,
  Kanayama, Saha, Sugawara, Sato, Higemoto, Koda, Ohishi, Nishiyama
  et~al.}}]{Aoki03}
\bibinfo{author}{\bibfnamefont{Y.}~\bibnamefont{Aoki}},
  \bibinfo{author}{\bibfnamefont{A.}~\bibnamefont{Tsuchiya}},
  \bibinfo{author}{\bibfnamefont{T.}~\bibnamefont{Kanayama}},
  \bibinfo{author}{\bibfnamefont{S.~R.} \bibnamefont{Saha}},
  \bibinfo{author}{\bibfnamefont{H.}~\bibnamefont{Sugawara}},
  \bibinfo{author}{\bibfnamefont{H.}~\bibnamefont{Sato}},
  \bibinfo{author}{\bibfnamefont{W.}~\bibnamefont{Higemoto}},
  \bibinfo{author}{\bibfnamefont{A.}~\bibnamefont{Koda}},
  \bibinfo{author}{\bibfnamefont{K.}~\bibnamefont{Ohishi}},
  \bibinfo{author}{\bibfnamefont{K.}~\bibnamefont{Nishiyama}},
  \bibnamefont{et~al.}, \bibinfo{journal}{Phys. Rev. Lett.}
  \textbf{\bibinfo{volume}{91}}, \bibinfo{pages}{067003}
  (\bibinfo{year}{2003}).

\bibitem[{\citenamefont{Luke et~al.}(1998)\citenamefont{Luke, Fudamoto, Kojima,
  Larkin, Merrin, Nachumi, Uemura, Maeno, Mao, Mori et~al.}}]{Luke98}
\bibinfo{author}{\bibfnamefont{G.~M.} \bibnamefont{Luke}},
  \bibinfo{author}{\bibfnamefont{Y.}~\bibnamefont{Fudamoto}},
  \bibinfo{author}{\bibfnamefont{K.~M.} \bibnamefont{Kojima}},
  \bibinfo{author}{\bibfnamefont{M.~I.} \bibnamefont{Larkin}},
  \bibinfo{author}{\bibfnamefont{J.}~\bibnamefont{Merrin}},
  \bibinfo{author}{\bibfnamefont{B.}~\bibnamefont{Nachumi}},
  \bibinfo{author}{\bibfnamefont{Y.~J.} \bibnamefont{Uemura}},
  \bibinfo{author}{\bibfnamefont{Y.}~\bibnamefont{Maeno}},
  \bibinfo{author}{\bibfnamefont{Z.~Q.} \bibnamefont{Mao}},
  \bibinfo{author}{\bibfnamefont{Y.}~\bibnamefont{Mori}}, \bibnamefont{et~al.},
  \bibinfo{journal}{Nature} \textbf{\bibinfo{volume}{394}},
  \bibinfo{pages}{558} (\bibinfo{year}{1998}).

\bibitem[{\citenamefont{Xia et~al.}(2006)\citenamefont{Xia, Maeno, Beyersdorf,
  Fejer, and Kapitulnik}}]{Xia06}
\bibinfo{author}{\bibfnamefont{J.}~\bibnamefont{Xia}},
  \bibinfo{author}{\bibfnamefont{Y.}~\bibnamefont{Maeno}},
  \bibinfo{author}{\bibfnamefont{P.}~\bibnamefont{Beyersdorf}},
  \bibinfo{author}{\bibfnamefont{M.}~\bibnamefont{Fejer}}, \bibnamefont{and}
  \bibinfo{author}{\bibfnamefont{A.}~\bibnamefont{Kapitulnik}},
  \bibinfo{journal}{Phys. Rev. Lett.} \textbf{\bibinfo{volume}{97}},
  \bibinfo{pages}{167002} (\bibinfo{year}{2006}).

\bibitem[{\citenamefont{Luke et~al.}(1993)\citenamefont{Luke, Keren, Le, Wu,
  Uemura, Bonn, Taillefer, and Garrett}}]{Luke93}
\bibinfo{author}{\bibfnamefont{G.~M.} \bibnamefont{Luke}},
  \bibinfo{author}{\bibfnamefont{A.}~\bibnamefont{Keren}},
  \bibinfo{author}{\bibfnamefont{L.~P.} \bibnamefont{Le}},
  \bibinfo{author}{\bibfnamefont{W.~D.} \bibnamefont{Wu}},
  \bibinfo{author}{\bibfnamefont{Y.~J.} \bibnamefont{Uemura}},
  \bibinfo{author}{\bibfnamefont{D.~A.} \bibnamefont{Bonn}},
  \bibinfo{author}{\bibfnamefont{L.}~\bibnamefont{Taillefer}},
  \bibnamefont{and} \bibinfo{author}{\bibfnamefont{J.~D.}
  \bibnamefont{Garrett}}, \bibinfo{journal}{Phys. Rev. Lett.}
  \textbf{\bibinfo{volume}{71}}, \bibinfo{pages}{1466} (\bibinfo{year}{1993}).

\bibitem[{\citenamefont{de~Reotier et~al.}(1995)\citenamefont{de~Reotier,
  Huxley, Yaouanc, Flouquet, Bonville, Impert, Pari, Gubbens, and
  Mulders}}]{deReotier95}
\bibinfo{author}{\bibfnamefont{P.~D.} \bibnamefont{de~Reotier}},
  \bibinfo{author}{\bibfnamefont{A.}~\bibnamefont{Huxley}},
  \bibinfo{author}{\bibfnamefont{A.}~\bibnamefont{Yaouanc}},
  \bibinfo{author}{\bibfnamefont{J.}~\bibnamefont{Flouquet}},
  \bibinfo{author}{\bibfnamefont{P.}~\bibnamefont{Bonville}},
  \bibinfo{author}{\bibfnamefont{P.}~\bibnamefont{Impert}},
  \bibinfo{author}{\bibfnamefont{P.}~\bibnamefont{Pari}},
  \bibinfo{author}{\bibfnamefont{P.~C.~M.} \bibnamefont{Gubbens}},
  \bibnamefont{and} \bibinfo{author}{\bibfnamefont{A.~M.}
  \bibnamefont{Mulders}}, \bibinfo{journal}{Phys. Lett. A}
  \textbf{\bibinfo{volume}{205}}, \bibinfo{pages}{239} (\bibinfo{year}{1995}).

\bibitem[{\citenamefont{Higemoto et~al.}(2000)\citenamefont{Higemoto, Satoh,
  Nishida, Koda, Nagamine, Haga, Yamamoto, Kimura, and Onuki}}]{Higemoto00}
\bibinfo{author}{\bibfnamefont{W.}~\bibnamefont{Higemoto}},
  \bibinfo{author}{\bibfnamefont{K.}~\bibnamefont{Satoh}},
  \bibinfo{author}{\bibfnamefont{N.}~\bibnamefont{Nishida}},
  \bibinfo{author}{\bibfnamefont{A.}~\bibnamefont{Koda}},
  \bibinfo{author}{\bibfnamefont{K.}~\bibnamefont{Nagamine}},
  \bibinfo{author}{\bibfnamefont{Y.}~\bibnamefont{Haga}},
  \bibinfo{author}{\bibfnamefont{E.}~\bibnamefont{Yamamoto}},
  \bibinfo{author}{\bibfnamefont{N.}~\bibnamefont{Kimura}}, \bibnamefont{and}
  \bibinfo{author}{\bibfnamefont{Y.}~\bibnamefont{Onuki}},
  \bibinfo{journal}{Physica B} \textbf{\bibinfo{volume}{281}},
  \bibinfo{pages}{984} (\bibinfo{year}{2000}).

\bibitem[{\citenamefont{Heffner et~al.}(1990)\citenamefont{Heffner, Smith,
  Willis, Birrer, Baines, Gygax, Hitti, Lippelt, Ott, Schenck
  et~al.}}]{Heffner90}
\bibinfo{author}{\bibfnamefont{R.~H.} \bibnamefont{Heffner}},
  \bibinfo{author}{\bibfnamefont{J.~L.} \bibnamefont{Smith}},
  \bibinfo{author}{\bibfnamefont{J.~O.} \bibnamefont{Willis}},
  \bibinfo{author}{\bibfnamefont{P.}~\bibnamefont{Birrer}},
  \bibinfo{author}{\bibfnamefont{C.}~\bibnamefont{Baines}},
  \bibinfo{author}{\bibfnamefont{F.~N.} \bibnamefont{Gygax}},
  \bibinfo{author}{\bibfnamefont{B.}~\bibnamefont{Hitti}},
  \bibinfo{author}{\bibfnamefont{E.}~\bibnamefont{Lippelt}},
  \bibinfo{author}{\bibfnamefont{H.~R.} \bibnamefont{Ott}},
  \bibinfo{author}{\bibfnamefont{A.}~\bibnamefont{Schenck}},
  \bibnamefont{et~al.}, \bibinfo{journal}{Phys. Rev. Lett.}
  \textbf{\bibinfo{volume}{65}}, \bibinfo{pages}{2816} (\bibinfo{year}{1990}).

\bibitem[{\citenamefont{Maisuradze et~al.}(2010)\citenamefont{Maisuradze,
  Schnelle, Khasanov, Gumeniuk, Nicklas, Rosner, Leithe-Jasper, Grin, Amato,
  and Thalmeier}}]{Maisuradze10}
\bibinfo{author}{\bibfnamefont{A.}~\bibnamefont{Maisuradze}},
  \bibinfo{author}{\bibfnamefont{W.}~\bibnamefont{Schnelle}},
  \bibinfo{author}{\bibfnamefont{R.}~\bibnamefont{Khasanov}},
  \bibinfo{author}{\bibfnamefont{R.}~\bibnamefont{Gumeniuk}},
  \bibinfo{author}{\bibfnamefont{M.}~\bibnamefont{Nicklas}},
  \bibinfo{author}{\bibfnamefont{H.}~\bibnamefont{Rosner}},
  \bibinfo{author}{\bibfnamefont{A.}~\bibnamefont{Leithe-Jasper}},
  \bibinfo{author}{\bibfnamefont{Y.}~\bibnamefont{Grin}},
  \bibinfo{author}{\bibfnamefont{A.}~\bibnamefont{Amato}}, \bibnamefont{and}
  \bibinfo{author}{\bibfnamefont{P.}~\bibnamefont{Thalmeier}},
  \bibinfo{journal}{Phys. Rev. B} \textbf{\bibinfo{volume}{82}},
  \bibinfo{pages}{024524} (\bibinfo{year}{2010}).

\bibitem[{\citenamefont{Shu et~al.}(2011)\citenamefont{Shu, Aoki, Hillier,
  Ohishi, Ishida, Kadono, Koda, Bernal, MacLaughlin, Tunashima et~al.}}]{Shu11}
\bibinfo{author}{\bibfnamefont{L.}~\bibnamefont{Shu}},
  \bibinfo{author}{\bibfnamefont{W.~H.~Y.} \bibnamefont{Aoki}},
  \bibinfo{author}{\bibfnamefont{A.~D.} \bibnamefont{Hillier}},
  \bibinfo{author}{\bibfnamefont{K.}~\bibnamefont{Ohishi}},
  \bibinfo{author}{\bibfnamefont{K.}~\bibnamefont{Ishida}},
  \bibinfo{author}{\bibfnamefont{R.}~\bibnamefont{Kadono}},
  \bibinfo{author}{\bibfnamefont{A.}~\bibnamefont{Koda}},
  \bibinfo{author}{\bibfnamefont{O.~O.} \bibnamefont{Bernal}},
  \bibinfo{author}{\bibfnamefont{D.~E.} \bibnamefont{MacLaughlin}},
  \bibinfo{author}{\bibfnamefont{Y.}~\bibnamefont{Tunashima}},
  \bibnamefont{et~al.}, \bibinfo{journal}{submitted to Phys. Rev. Lett.}
  (\bibinfo{year}{2011}).

\bibitem[{\citenamefont{Adroja et~al.}(2005)\citenamefont{Adroja, Hillier,
  Park, Goremychkin, Takeda, Osborn, Rainford, and Ibberson}}]{Adroja05}
\bibinfo{author}{\bibfnamefont{D.~T.} \bibnamefont{Adroja}},
  \bibinfo{author}{\bibfnamefont{A.~D.} \bibnamefont{Hillier}},
  \bibinfo{author}{\bibfnamefont{J.~G.} \bibnamefont{Park}},
  \bibinfo{author}{\bibfnamefont{E.~A.} \bibnamefont{Goremychkin}},
  \bibinfo{author}{\bibfnamefont{N.}~\bibnamefont{Takeda}},
  \bibinfo{author}{\bibfnamefont{R.}~\bibnamefont{Osborn}},
  \bibinfo{author}{\bibfnamefont{B.~D.} \bibnamefont{Rainford}},
  \bibnamefont{and} \bibinfo{author}{\bibfnamefont{R.~M.}
  \bibnamefont{Ibberson}}, \bibinfo{journal}{Phys. Rev. B}
  \textbf{\bibinfo{volume}{72}}, \bibinfo{pages}{184503}
  (\bibinfo{year}{2005}).

\bibitem[{\citenamefont{Anand et~al.}(2011)\citenamefont{Anand, Hillier,
  Adroja, Strydom, Michor, McEwen, and Rainford}}]{Anand11}
\bibinfo{author}{\bibfnamefont{V.}~\bibnamefont{Anand}},
  \bibinfo{author}{\bibfnamefont{A.~D.} \bibnamefont{Hillier}},
  \bibinfo{author}{\bibfnamefont{D.~T.} \bibnamefont{Adroja}},
  \bibinfo{author}{\bibfnamefont{A.~M.} \bibnamefont{Strydom}},
  \bibinfo{author}{\bibfnamefont{H.}~\bibnamefont{Michor}},
  \bibinfo{author}{\bibfnamefont{K.~A.} \bibnamefont{McEwen}},
  \bibnamefont{and} \bibinfo{author}{\bibfnamefont{B.~D.}
  \bibnamefont{Rainford}}, \bibinfo{journal}{Phys. Rev. B}
  \textbf{\bibinfo{volume}{83}}, \bibinfo{pages}{064552}
  (\bibinfo{year}{2011}).

\bibitem[{\citenamefont{Tran et~al.}(2010)\citenamefont{Tran, Hillier, and
  Adroja}}]{Tran10}
\bibinfo{author}{\bibfnamefont{V.~H.} \bibnamefont{Tran}},
  \bibinfo{author}{\bibfnamefont{A.~D.} \bibnamefont{Hillier}},
  \bibnamefont{and} \bibinfo{author}{\bibfnamefont{D.~T.}
  \bibnamefont{Adroja}}, \bibinfo{journal}{J. Phys. Cond. Matter}
  \textbf{\bibinfo{volume}{22}}, \bibinfo{pages}{505701}
  (\bibinfo{year}{2010}).

\bibitem[{\citenamefont{Hillier et~al.}(2011)\citenamefont{Hillier, Parzyk, and
  Paul}}]{Hillier11}
\bibinfo{author}{\bibfnamefont{A.~D.} \bibnamefont{Hillier}},
  \bibinfo{author}{\bibfnamefont{N.}~\bibnamefont{Parzyk}}, \bibnamefont{and}
  \bibinfo{author}{\bibfnamefont{D.~M.} \bibnamefont{Paul}},
  \bibinfo{journal}{Phys. Rev. B submitted}  (\bibinfo{year}{2011}).

\bibitem[{\citenamefont{Mineev and Samokhin}(1999)}]{Mineev99}
\bibinfo{author}{\bibfnamefont{V.~P.} \bibnamefont{Mineev}} \bibnamefont{and}
  \bibinfo{author}{\bibfnamefont{K.~K.} \bibnamefont{Samokhin}},
  \emph{\bibinfo{title}{Introduction to Unconventional Superconductivity}}
  (\bibinfo{publisher}{Gordon and Breach}, \bibinfo{year}{1999}).

\bibitem[{\citenamefont{Bonalde et~al.}(2011)\citenamefont{Bonalde, Ribeiro,
  Syu, Sung, and Lee}}]{Bonalde11}
\bibinfo{author}{\bibfnamefont{I.}~\bibnamefont{Bonalde}},
  \bibinfo{author}{\bibfnamefont{R.~L.} \bibnamefont{Ribeiro}},
  \bibinfo{author}{\bibfnamefont{K.~J.} \bibnamefont{Syu}},
  \bibinfo{author}{\bibfnamefont{H.~H.} \bibnamefont{Sung}}, \bibnamefont{and}
  \bibinfo{author}{\bibfnamefont{W.~H.} \bibnamefont{Lee}},
  \bibinfo{journal}{New Journal of Physics} \textbf{\bibinfo{volume}{13}},
  \bibinfo{pages}{123022} (\bibinfo{year}{2011}).

\bibitem[{\citenamefont{Zeng and Lee}(2002)}]{Zeng02}
\bibinfo{author}{\bibfnamefont{N.~L.} \bibnamefont{Zeng}} \bibnamefont{and}
  \bibinfo{author}{\bibfnamefont{W.~H.} \bibnamefont{Lee}},
  \bibinfo{journal}{Phys. Rev. B} \textbf{\bibinfo{volume}{66}},
  \bibinfo{pages}{092503} (\bibinfo{year}{2002}).

\bibitem[{\citenamefont{Nishizaki et~al.}(2000)\citenamefont{Nishizaki, Maeno,
  and Mao}}]{Nishizaki00}
\bibinfo{author}{\bibfnamefont{S.}~\bibnamefont{Nishizaki}},
  \bibinfo{author}{\bibfnamefont{Y.}~\bibnamefont{Maeno}}, \bibnamefont{and}
  \bibinfo{author}{\bibfnamefont{Z.}~\bibnamefont{Mao}}, \bibinfo{journal}{J.
  Phys. Soc. Japan} \textbf{\bibinfo{volume}{69}}, \bibinfo{pages}{572}
  (\bibinfo{year}{2000}).

\bibitem[{\citenamefont{Pecharsky et~al.}(1998)\citenamefont{Pecharsky, Miller,
  and K.~A.~Gschneider}}]{Pecharsky98}
\bibinfo{author}{\bibfnamefont{V.~K.} \bibnamefont{Pecharsky}},
  \bibinfo{author}{\bibfnamefont{L.~L.} \bibnamefont{Miller}},
  \bibnamefont{and}
  \bibinfo{author}{\bibfnamefont{J.}~\bibnamefont{K.~A.~Gschneider}},
  \bibinfo{journal}{Phys. Rev. B} \textbf{\bibinfo{volume}{58}},
  \bibinfo{pages}{497} (\bibinfo{year}{1998}).

\bibitem[{\citenamefont{Gumeniuk et~al.}(2008)\citenamefont{Gumeniuk, Schnelle,
  Rosner, Nicklas, Leithe-Jasper, and Grin}}]{Gumeniuk08}
\bibinfo{author}{\bibfnamefont{R.}~\bibnamefont{Gumeniuk}},
  \bibinfo{author}{\bibfnamefont{W.}~\bibnamefont{Schnelle}},
  \bibinfo{author}{\bibfnamefont{H.}~\bibnamefont{Rosner}},
  \bibinfo{author}{\bibfnamefont{M.}~\bibnamefont{Nicklas}},
  \bibinfo{author}{\bibfnamefont{A.}~\bibnamefont{Leithe-Jasper}},
  \bibnamefont{and} \bibinfo{author}{\bibfnamefont{Y.}~\bibnamefont{Grin}},
  \bibinfo{journal}{Phys. Rev. Lett.} \textbf{\bibinfo{volume}{100}},
  \bibinfo{pages}{017002} (\bibinfo{year}{2008}).

\bibitem[{\citenamefont{Lee et~al.}(1999)\citenamefont{Lee, Kilcoyne, and
  Cywinski}}]{blackbook}
\bibinfo{author}{\bibfnamefont{S.~L.} \bibnamefont{Lee}},
  \bibinfo{author}{\bibfnamefont{S.~H.} \bibnamefont{Kilcoyne}},
  \bibnamefont{and} \bibinfo{author}{\bibfnamefont{R.}~\bibnamefont{Cywinski}},
  \emph{\bibinfo{title}{Muon Science: Muons in Physics, Chemistry and
  Materials}} (\bibinfo{publisher}{SUSSP Publications and IOP Publishing},
  \bibinfo{year}{1999}).

\bibitem[{\citenamefont{Yaouanc and de~Reotier}(2011)}]{Yaouanc}
\bibinfo{author}{\bibfnamefont{A.}~\bibnamefont{Yaouanc}} \bibnamefont{and}
  \bibinfo{author}{\bibfnamefont{P.~D.} \bibnamefont{de~Reotier}},
  \emph{\bibinfo{title}{Muon Spin Rotation, Relaxation, and Resonance}}
  (\bibinfo{publisher}{Oxford University Press}, \bibinfo{year}{2011}).

\bibitem[{\citenamefont{Schenck}(1985)}]{Schenck}
\bibinfo{author}{\bibfnamefont{A.}~\bibnamefont{Schenck}},
  \emph{\bibinfo{title}{Muon Spin Rotation Spectroscopy Principles and
  Applications in Solid State Physics}} (\bibinfo{publisher}{Taylor and
  Francis}, \bibinfo{year}{1985}).

\bibitem[{\citenamefont{Hillier and Cywinski}(1997)}]{Hillier97}
\bibinfo{author}{\bibfnamefont{A.~D.} \bibnamefont{Hillier}} \bibnamefont{and}
  \bibinfo{author}{\bibfnamefont{R.}~\bibnamefont{Cywinski}},
  \bibinfo{journal}{Applied Magnetic Resonance} \textbf{\bibinfo{volume}{13}},
  \bibinfo{pages}{95} (\bibinfo{year}{1997}).

\bibitem[{\citenamefont{Hayano et~al.}(1979)\citenamefont{Hayano, Uemura,
  Imazato, Nishida, Yamazaki, and Kubo}}]{Hayano79}
\bibinfo{author}{\bibfnamefont{R.~S.} \bibnamefont{Hayano}},
  \bibinfo{author}{\bibfnamefont{Y.~J.} \bibnamefont{Uemura}},
  \bibinfo{author}{\bibfnamefont{J.}~\bibnamefont{Imazato}},
  \bibinfo{author}{\bibfnamefont{N.}~\bibnamefont{Nishida}},
  \bibinfo{author}{\bibfnamefont{T.}~\bibnamefont{Yamazaki}}, \bibnamefont{and}
  \bibinfo{author}{\bibfnamefont{R.}~\bibnamefont{Kubo}},
  \bibinfo{journal}{Phys. Rev. B} \textbf{\bibinfo{volume}{20}},
  \bibinfo{pages}{850} (\bibinfo{year}{1979}).

\bibitem[{\citenamefont{Carrington and Manzano}(2003)}]{Carrington03}
\bibinfo{author}{\bibfnamefont{A.}~\bibnamefont{Carrington}} \bibnamefont{and}
  \bibinfo{author}{\bibfnamefont{F.}~\bibnamefont{Manzano}},
  \bibinfo{journal}{Physica C} \textbf{\bibinfo{volume}{385}},
  \bibinfo{pages}{205} (\bibinfo{year}{2003}).

\end{thebibliography}
\end{document}